\documentclass[11pt, prd,preprintnumbers, amsmath,amssymb, superscriptaddress, nofootinbib, hyperref]{revtex4}

\pdfoutput=1
\usepackage{latexsym}
\usepackage{wasysym}
\usepackage{amssymb}
\usepackage{cancel}
\usepackage{epsfig,amsmath,graphics}
\usepackage{epstopdf}
\usepackage{verbatim}
\usepackage[hypertex]{hyperref}
\usepackage{graphicx}% Include figure files
\usepackage{subfigure}
\usepackage[version=3]{mhchem}

\newcommand{\ChiOne}{\chi_{g}}
\newcommand{\ChiTwo}{\chi_{e}}
\newcommand{\Splitting}{\delta}
\newcommand{\ChiMass}{m_{\chi}}

\newcommand{\tmop}[1]{\ensuremath{\operatorname{#1}}}

\def\r{\right)}
\def\l{\left(}

\begin{document}

\title{Luminous Dark Matter}

\date{\today}% It is always \today, today,
             %  but any date may be explicitly specified
             
\author{Brian Feldstein}
%\email{pwgraham@stanford.edu}
\affiliation{Department of Physics, Boston University, Boston, MA 02215}
             
 \author{Peter W. Graham}
%\email{pwgraham@stanford.edu}
\affiliation{Department of Physics, Stanford University, Stanford, California 94305}

\author{Surjeet Rajendran}
%\email{surjeet@stanford.edu}
\affiliation{Center for Theoretical Physics, Laboratory for Nuclear Science and Department of Physics, Massachusetts Institute of Technology, Cambridge, MA 02139, USA}

\preprint{MIT-CTP 4172}

\begin{abstract}
We propose a dark matter model in which the signal in direct detection experiments arises from electromagnetic, not nuclear, energy deposition.  This can provide a novel explanation for DAMA while avoiding many direct detection constraints.  The dark matter state is taken nearly degenerate with another state.  These states are naturally connected by a dipole moment operator, which can give both the dominant scattering and decay modes between the two states.  The signal at DAMA then arises from dark matter scattering in the Earth into the excited state and decaying back to the ground state through emission of a single photon in the detector.  This model has unique signatures in direct detection experiments.  The density and chemical composition of the detector is irrelevant, only the total volume affects the event rate.  In addition, the spectrum is a monoenergetic line, which can fit the DAMA signal well.  This model is readily testable at experiments such as CDMS and XENON100 if they analyze their low-energy, electronic recoil events.
\end{abstract}

%\pacs{}% PACS, the Physics and Astronomy
                             % Classification Scheme.
%\keywords{Suggested keywords}%Use showkeys class option if keyword
                              %display desired
\maketitle
\tableofcontents

\section{Introduction and Overview}
\label{Sec:Intro}

The signal in DAMA is now highly statistically significant \cite{Bernabei:2008yi, Bernabei:2010mq}.  However, the interpretation is still unclear.  The constraints from many other direct detection experiments, including CDMS \cite{Ahmed:2009zw} and XENON100 \cite{Aprile:2010um}, rule out the simplest interpretation in terms of a WIMP elastically scattering off nuclei.  Of course, comparing these different experimental constraints is highly model-dependent and indeed several models have been put forward to explain the positive signal in DAMA while avoiding other direct detection constraints including inelastic dark matter (iDM) \cite{TuckerSmith:2001hy}, exothermic dark matter (exoDM) \cite{Graham:2010ca}, form factor dark matter \cite{Feldstein:2009np, Feldstein:2009tr, Chang:2009yt}, resonant dark matter (rDM) \cite{Bai:2009cd}, and light dark matter \cite{Bottino:2003cz, Petriello:2008jj, Savage:2008er}.  All these models rely on nuclear scattering in DAMA to explain their signal.  We wish to propose a different explanation.

Many models beyond the standard model have several states, of which one (or more) is stable and thus makes up the dark matter of our universe.  It is natural for these states to be mixed by a dipole moment operator.  Such operators are induced by loops in many theories including supersymmetry.  Although the dipole moment operator may be avoided, it seems generic to consider the possibility that it exists at some level.  We will then make one further assumption, that the mass splitting between the dark matter state and another state in the spectrum is small, $\sim$ keV.  Many models that explain DAMA make use of such  small mass splittings.  We will argue that these assumptions, two nearly degenerate states connected by a dipole moment operator, can explain DAMA by electronic instead of nuclear scattering events, thus avoiding the constraints from most other direct detection experiments.  DAMA is one of the only direct detection experiments that does not attempt to veto electronic events.

Because the splitting is small and the dipole moment is a higher dimension operator suppressed, for example by $\sim$ TeV, the excited state of dark matter can actually live for a significant amount of time.  This means that, just due to the dipole moment operator, dark matter will upscatter into the excited state in the Earth and then decay back to the ground state a significant distance away.  These decays happen by single photon emission and thus, when they happen in a detector will appear as an electromagnetic, not nuclear, scattering.  As explained more completely in Section \ref{Sec:DipolarDarkMatter}, the decay rate per unit volume in the Earth is naturally close to the original nuclear scattering rate and thus of the right order to explain DAMA.  Note that this explanation for DAMA is quite different from recently proposed models which involve a dipole moment induced nuclear scattering \cite{Chang:2010en, Fitzpatrick:2010br, Banks:2010eh}.

Generally, electronic scattering has not worked as an explanation for DAMA due to the large form factor that suppresses the scattering of a heavy WIMP off a light electron \cite{Kopp:2009et}.  Alternatively, absorbing light axion-like dark matter leads to a very low annual modulation \cite{Pospelov:2008jk}.  Our model avoids such problems because the dark matter does not actually scatter off electrons.  Instead it produces a photon which then easily dumps all its energy in the detector.  Because the dark matter is slow, the photons produced are monoenergetic and appear as a line spectrum in a detector.  Interestingly, a line, smeared by DAMA's energy resolution, fits the DAMA signal well.  In fact, more generally than explaining DAMA, this model can produce electron-recoil events instead of nuclear-recoil events without a large form factor suppression.

\section{Luminous Dark Matter }
\label{Sec:DipolarDarkMatter}

In its most general form, the scenario we are considering requires two components:
First, we require an interaction to mediate an inelastic scattering of the
dark matter particle within the Earth, up to a higher energy state, and second, we
require an interaction leading to a decay of this higher energy state into a
set of products which includes photons. \ Of course, it is also necessary to
ensure that the interactions available do not also lead to unacceptably fast
decays of the original dark matter particle.

Here we will concentrate on a simple scenario in which
all of the requirements are satisfied by a single dark sector interaction
term. \ In particular, we will take an interaction which results from the
magnetic dipole moment operator
\begin{equation}
  \mathcal{L}_M = \frac{i}{4 \Lambda} \bar{\chi}_g \sigma^{\mu \nu} \chi_e
  F_{\mu \nu} + h.c., \label{MDM}
\end{equation}
where here $\ChiOne$ and $\ChiTwo$ are two distinct fermions, which we will take
to be Majorana for simplicity. \ We will discuss a straightforward mechanism
leading to such an interaction in section \ref{Sec:IDMOperator}.{\footnote{We assume that CP is a good enough symmetry in the dark sector so as to sufficiently suppress a
possible electric dipole moment operator, which would otherwise be the
dominant channel for scattering due to a $1 / v^2$ enhancement.}}

We will take the dark matter particle to be $\chi_g$, and to have
a mass $m_{\chi} \sim 1 \tmop{GeV}$, while $\chi_e$ will have a mass
which is higher than this by a splitting $\delta$ of size $3.3 \tmop{keV}$,
chosen to fit the observed DAMA spectrum. \
The light dark matter masses we employ will be heavy enough to allow for
sufficient kinetic energy for upscattering, while also being light enough to
allow the required cross sections to evade current direct detection limits;
the nuclear recoil events simply deposit too little energy to have been seen
by present dark matter searches. \ The photon decay products, on the other
hand, lead to various constraints on the model which will be discussed in
detail in section \ref{Sec:Constraints}.

\begin{figure}
\begin{center}
\includegraphics[width = 5.5 in]{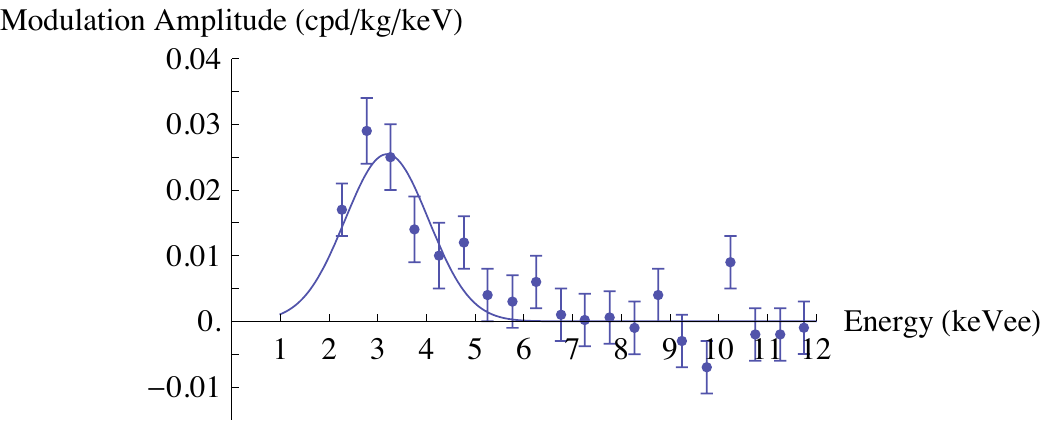}
\caption{ \label{Fig:DAMA35} A comparison of the predicted and observed
  spectra at DAMA for a line at 3.3 keV using the assumptions defined in the
  text. The dots represent the DAMA data points, while the line is an example
of a fit from our model.}    
\end{center}
\end{figure}

After the dark matter particle has upscattered in the Earth to the
higher energy state, the resulting particle may travel a great distance before
ultimately decaying. \ Note that there is no need for both processes to occur
within the DAMA detector. \ Indeed, if the typical travel distance before the
decay is less than roughly the Earth's radius, then, with the cross-section
fixed, the signal at DAMA is actually independent of the decay rate. \ This is
easy to understand, since the upscatter and decay events are then spread
roughly evenly throughout the Earth, and thus automatically have comparable
rates per unit volume, on average. \ Indeed, this regime will apply throughout
the viable parameter space of the model. \ Note that a $3.3 \tmop{keV}$ line
gives an excellent fit to the spectrum observed at DAMA, due to the energy
resolution of the detectors of about $.8 \tmop{keV}$ at this energy scale \cite{Bernabei:2008yh}. \ A
comparison between the observed and predicted spectra is shown in figure \ref{Fig:DAMA35} .

Before proceeding to the calculation of the event rate at DAMA,
let us briefly consider some of the key similarities and differences between this
scenario, and the usual case of inelastic dark matter.  In the inelastic dark matter scenario, interactions mediating
elastic scattering are required to be significantly suppressed compared to the
dominant inelastic interactions. \ If this were not the case, the elastic
interactions would lead to unacceptably large signals at direct detection
experiments. \ In the present model, however, the constraints on the elastic
scattering cross-section are much weaker; indeed, there is no problem having
elastic cross-sections of the same order as the inelastic ones {\footnote{Note
on the other hand that a magnetic dipole interaction could not mediate elastic
scattering of our Majorana dark matter particle, since such an interaction
would vanish by fermion anti-commutation.}}. \ An important feature which our
scenario has in common with inelastic dark matter, on the other hand, lies in
the possibility for the existence of a boost to the annual modulation
fraction. \ At the small masses we consider, only dark matter particles on the
rapidly falling tail of the halo velocity distribution have sufficient kinetic
energy to upscatter, leading to the boost. \ For standard WIMPS, this would
lead to an unsuitably sharp spectrum at DAMA, but in both our scenario and
iDM, this problem is solved by the details of the mechanism; in iDM the
spectrum is fixed by the cutoff in available low energy phase-space, while in
our model the spectrum is fixed due to its origin in the monochromatic photon
decay. \ This boost to the annual modulation fraction is needed due to the
bounds on a purely electromagnetic signal at the XENON100 experiment \cite{Aprile:2010um} (and to a lesser extent, the CDMS experiment). \ As will be discussed further in
section \ref{Sec:Constraints}, we require modulation fractions larger than around $50\%$, and this will constrain from above the masses for which our scenario is viable.

We now turn to a calculation of the event rate at DAMA. \ In
principle, this calculation is made quite complicated by the angular
dependence of the inelastic dipole scattering cross-section. \ Events leading
to a signal at DAMA must involve dark matter particles upscattered within the
Earth towards the direction of the experiment. The probability for this to
occur depends on the relative locations in the Earth of the initial
upscattering event and the DAMA detector, as well as on the orientation of the
incoming dark matter velocity. \ A complete calculation of the predicted rate
at DAMA is therefore fairly complicated, and in practice is rather
time-consuming to perform numerically. \ As a result we will content ourselves
with calculating the expected event rate -and required cross-sections- to
within about a factor of $\sim 4$, and will make a number of simplifying
approximations. \ The number of events we predict to have been seen at
XENON100 follows directly from the DAMA event rate, and the given modulation
fraction. \ Since the modulation fraction is essentially fixed by the dark
matter mass and not the overall cross-section, the severity of the XENON100
constraint is essentially independent of any uncertainty introduced through
our approximations. \ The other main constraint on our model comes from
the SWIFT x-ray satellite \cite{Moretti:2008hs}, and will be discussed further in section
\ref{Sec:Constraints}.  Generally, the approximations we make should tend to
underestimate the required cross-sections and decay rates slightly, thus making the SWIFT
constraint appear a bit stronger than it otherwise might;  faster decay rates
result in fewer $\chi_e$ particles reaching the height of the satellite.
Note, however, that uncertainty in the composition of the Earth could affect
the SWIFT constraint in either direction.

The approximations we will make are as follows:

1. We will take the Earth to have radius $r_{\oplus} = 6.4 \times
10^6\tmop{m}$, and to consist of a uniform crust of density $2.9 \tmop{g} / \tmop{cm}^3$
and depth $40 \tmop{km}$, and a mantle region with density increasing linearly
with depth from $3.3 \tmop{g} / \tmop{cm}^3$ to $5.7 \text{g} / \tmop{cm}^3$. \ The Earth's
core will be taken to be a uniform sphere with radius $3500 \tmop{km}$ and
density $10.5 \tmop{g} / \tmop{cm}^3$. \ Note however that for the viable parameter
space of our model, the typical distances travelled by upscattered dark matter
particles before they decay is on the order of $100 - 1000 \tmop{km}$, and thus
upscattering in the core does not actually contribute to the signal. \ The
elemental abundances we take for each section of the Earth are given in Table
\ref{elements}.{\footnote{We ignore various elements with small
abundances.}} \ It should be kept in mind that we do not actually know the
specific elemental abundances to be found in the vicinity of the DAMA
experiment, and so we simply take typical values {\footnote{Note that since
both DAMA and XENON100 are located in the Gran Sasso mine, the relative
signals seen by these experiments is not influenced by this approximation.}}.

\begin{table}[h]
  \begin{tabular}{|c|c|c|c|}
    Element & Crust Abundance by Mass & Mantle Abundance by Mass & Core
    Abundance by Mass\\
    \hline
    O & 46\% & 45\% & 0\%\\
    \hline
    Mg & 3\% & 23\% & 0\%\\
    \hline
    Al & 9\% & 2\% & 0\%\\
    \hline
    Si & 29\% & 22\% & 6\%\\
    \hline
    Ca & 4\% & 2\% & 0\%\\
    \hline
    Fe & 5\% & 6\% & 86\%\\
    \hline
  \end{tabular}
  \caption{\label{elements} The elemental abundances of the Earth assumed for our calculations, obtained from \cite{EarthAbundances}.  }
\end{table}

2. In determining the scattering cross-sections and kinematics for
upscattering, we will approximate all relevant elements in the Earth as being
infinitely heavy compared to the dark matter particle. \ More realistically,
scattering from lighter elements, especially oxygen, would be suppressed or
even kinematically forbidden. \ However, we estimate the error introduced in
this way to be less than about a factor of 2. \footnote{Indeed the changes that result from this approximation can be compensated for by making a small adjustment to the dark matter mass.} \ This will result in the
important simplification that the upscattered dark matter velocity will be
independent of the scattering angle.

3. We will work with upscattering cross-sections which have been
integrated over angles, rather than with the true angular dependent
differential cross-sections. \ We estimate that this approximation will
introduce an error of less than about a factor of 2. Note that the true cross-sections are in fact angular
independent at the threshold for upscattering. \ Beyond threshold, on the
other hand, they quickly become fairly forward peaked. \ The reason our
approximation is a good one, is that the falling halo velocity distribution
results in most upscattering events taking place quite close to threshold.

With these approximations in hand, calculating the event rate at
DAMA is straightforward. \ The total rate per unit detector mass is given by
\begin{equation}
  \mathcal{R}= \frac{\rho_{\tmop{DM}}}{m_{\chi} \, \rho_{\tmop{NaI}}} \sum_i
  \int d^3 \vec{v} f ( \vec{v}) \int_{\tmop{Earth}} d^3 \vec{r} n_i ( \vec{r})
  \sigma_i v\mathcal{P}( \vec{r}, v) . \label{rate}
\end{equation}
Here $f ( \vec{v})$ is the dark matter velocity distribution in the Earth
frame, which we will take to have the standard form
\begin{equation}
  f ( \vec{v}) = \frac{1}{(\pi \bar{v}^2)^{3 / 2}} e^{- ( \vec{v} +
  \vec{v}_e)^2 / \bar{v}^2} \Theta (v_{\tmop{esc}} - | \vec{v} + \vec{v}_e |),
  \label{f}
\end{equation}
with a cutoff in the galactic frame at $v_{\tmop{esc}} = 650 \tmop{km} / \text{s}$ \footnote{Taking $v_{\tmop{esc}} = 600 \tmop{km} / \text{s}$ would simply shift the required mass range slightly. }. \
We will take $\bar{v}$ to be $220 \tmop{km} / \text{s}$ and the galactic speed of
Earth, $v_e$, to be $232 \tmop{km} / \text{s} + 15 \cos \l 2 \pi (t - t_0) \r
\tmop{km} / \text{s}$, with $t_0 = \tmop{June} \tmop{2nd}$ and $t$ in years. \
$\rho_{\tmop{DM}}$ is the local dark matter density, which we will set to $.3
\tmop{GeV} / \tmop{cm}^3$, and $\rho_{\tmop{NaI}}$ is the sodium iodide
density of $3.67 \tmop{g} / \tmop{cm}^3$. \ $\sigma_i$ is the total cross-section to
upscatter from a given element $i$, with $i$ running over the rows of table
[\ref{elements}], and $n_i ( \vec{r})$ is the number density distribution in
the Earth for the given element. \ $\mathcal{P}( \vec{r}, v)$ is the
probability that an upscattering event in the Earth at position $\vec{r}$ with
incoming speed v leads to a decay within the DAMA dark matter detector,
divided by the detector volume. \ After upscattering, the excited state moves
with speed $v_f = \sqrt{v^2 - 2 \frac{\delta}{m_{\chi}}}$, so that we
have
\begin{equation}
  \mathcal{P}( \vec{r}, v) = \frac{1}{4 \pi ( \vec{r} -
  \vec{r}_{\tmop{DAMA}})^2} \frac{\Gamma}{v_f} e^{- \frac{\Gamma | \vec{r} -
  \vec{r}_{\tmop{DAMA}} |}{v_f}}, \label{P}
\end{equation}
with $r_{\tmop{DAMA}}$ being the position in the Earth of the detector.

We calculated the results for equation [\ref{rate}] numerically,
extracted the amplitude of the annual modulation, and can then compare with
the measured rate at DAMA.

\subsection{Results}

For the magnetic dipole operator [\ref{MDM}], the leading order
differential cross-section (in the center-of-mass frame) on a heavy element of
atomic number Z is
\begin{equation}
  \frac{d \sigma_M}{d \Omega} = \frac{e^2 \, Z^2 \,  [\delta^2 \, m_{\chi}^2 \,  -  \, 4 \, 
  \delta m_{\chi}^3 \,  v^2 \,  + \,  2 \, m_{\chi}^4 \,  v^4 \,  - \,  2 \, m_{\chi}^2 \,  v \,  v_f \, 
  (m_{\chi}^2 \,  v^2 \,  - \,  \delta m_{\chi}) \,  \cos \l \theta \r]}{16 \,  \pi^2 \, 
  \Lambda^2 \,  [\delta m_{\chi}\,  - \, m_{\chi}^2 \,  v^2 \,  + \, m_{\chi}^2 \,  v \, 
  v_f \,  \cos \l \theta \r ]^2}, \label{fullMDMsigma}
\end{equation}
which after integrating over angles becomes
\begin{equation}
  \sigma_M = \frac{e^2 Z^2 [2 (m_{\chi} v^2 - \delta)
  \tanh^{-1} ( \frac{m_{\chi} \, v \,  v_f}{m_{\chi} v^2 - \delta}) -
 m_{\chi} \, v \, v_f]}{4 \pi \Lambda^2m_{\chi} \, v \, v_f} .
  \label{sigmaMDM}
\end{equation}
Note that for all
elements of interest, and for $m_{\chi}$ in the desired range,
dipole-charge scattering always dominates over dipole-dipole scattering. \
Nuclear form factors are irrelevant due to the small momentum transfers
present in our scattering events.
The decay rate of the excited state $\ChiTwo$ is
\begin{equation}
  \Gamma = \frac{1}{\pi \Lambda^2} \delta^3 . \label{gamma}
\end{equation}

\begin{figure}
\begin{center}
\includegraphics[width = 4.1 in]{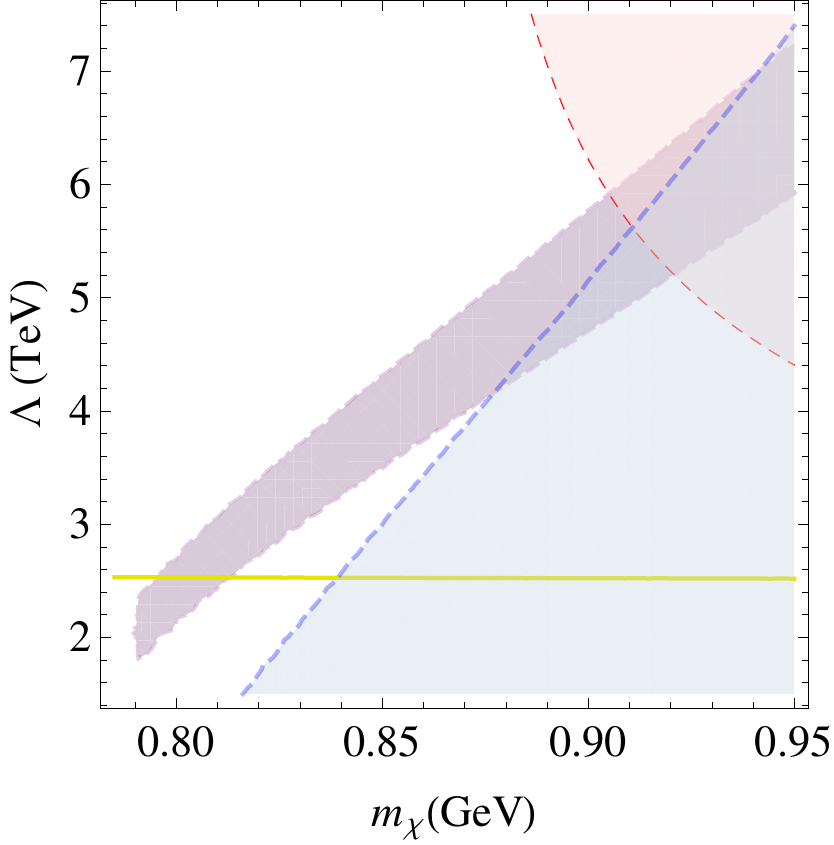}
\caption{ \label{Fig:LDMPlot}The parameter space available to fit the DAMA
  signal at 90 $\%$ confidence, using halo assumptions defined in the
  text and a $3.3 \tmop{keV}$ mass splitting. The vertical thickness of the purple region is essentially fixed by
  the measured DAMA event rate.  The blue dashed line shows the 90\% confidence XENON100 constraint,
  according to the assumptions defined in the text. The red line gives the
  x-ray background constraint from the SWIFT satellite.  The yellow line shows the
  parameter points which produce the correct thermal relic density.    }    
\end{center}
\end{figure}

The available parameter region in the $m_{\chi}$, $\Lambda$
plane is plotted in figure \ref{Fig:LDMPlot}. \ The purple region shows the parameter space
which reproduces the correct rate at DAMA at $90\%$ confidence, comparing with
the lowest 8 points in the DAMA energy spectrum, using a $3.3 \tmop{keV}$
mass splitting. Note that the vertical
thickness of this region is essentially set by the accuracy of the
measurement of the DAMA event rate.  \ The blue dashed line shows
the 90\% confidence XENON100 constraint, which we have obtained by comparing
the number of observed and predicted events within one standard deviation of
energy resolution from the predicted peak. \ We apply Poisson statistics to
the number of S1 photo-electrons produced in an event to obtain the energy
resolution (XENON100 claims a light yield of 2.2 photo-electrons/keV). We use
the energy dependent acceptance plotted in figure 3 of \cite{Aprile:2010um}, although we should
note that this acceptance is actually that for nuclear recoils (before cuts
from the S2 signal). \ No acceptance for electromagnetic events has been
published by the XENON100 collaboration at the present time, although the two
acceptances
are expected to be similar \cite{KaixuanNi}.  The red dashed line shows the
constraint from the measurements of the x-ray background by the SWIFT
satellite, and will be discussed further in section \ref{Sec:Constraints}.
\ The yellow line shows the parameter points which produce
the correct thermal relic density. \ It is an appealing feature of LDM that
the thermal relic density turns out to be roughly the correct size, simply
from the requirement of meeting all of the other experimental constraints. \
Note that the uncertainty at the factor of $\sim 4$ level in our cross-section
calculation will not spoil this statement; small changes to the assumed dark
matter halo velocity distribution could be made in order to maintain the
successful relic density if necessary.

Given the allowed range of values for $\Lambda$ shown in the figure, the
charge scattering cross section per proton in a given nucleus ranges
from about $10^{-38} \tmop{cm}^2$ to about $10^{-36} \tmop{cm}^2$. 

\subsection{Generating Inelastic Dipole Interactions}
\label{Sec:IDMOperator}

Constructing models which lead to interactions of the form
[\ref{MDM}] is straightforward. \ Essentially, we only need a few ingredients;
first, a Dirac fermion $\psi = \left(\begin{array}{c}
  \psi_{L_1}\\
  i \sigma_2 \psi_{L_2}^{\ast}
\end{array}\right)$ with a $\sim 1 \tmop{GeV}$ Dirac mass; second, a $3.3
\tmop{keV}$ Majorana mass for one of the two components of $\psi$, say
$\psi_{L_2}$; third, a magnetic dipole interaction for $\psi$. \ This dipole
interaction could arise, for example, from integrating out a heavy Dirac
fermion and a heavy scalar, each carrying hypercharge, and with a Yukawa
coupling to $\psi$. In general, magnetic dipole interactions are CP
conserving, and  may arise in many models
after integrating out heavy particles.\ In the present scenario, the mass
eigenstate fields are, at leading order, $\chi_{L_1} = \frac{i}{\sqrt{2}}
(\psi_{L_1} - \psi_{L_2})$ and $\chi_{L_2} = \frac{1}{\sqrt{2}} (\psi_{L_1} +
\psi_{L_2})$, with masses split by $3.3 \tmop{keV}$. \ Writing $\ChiOne =
\left(\begin{array}{c}
  \chi_{L_1}\\
  i \sigma_2 \chi_{L_1}^{\ast}
\end{array}\right) $ and $\ChiTwo = \left(\begin{array}{c}
  \chi_{L_2}\\
  i \sigma_2 \chi_{L_2}^{\ast}
\end{array}\right) $, and expanding out the original dipole interaction for
$\psi$, we obtain the inelastic dipole operator, [\ref{MDM}], as required.

\section{Constraints}
\label{Sec:Constraints}
In the scenario described in section \ref{Sec:DipolarDarkMatter}, the dark matter consists of a single state $\ChiOne$ with a mass $\ChiMass$. There is however another state $\ChiTwo$, which is slightly heavier than $\ChiOne$ by an amount $\Splitting \sim 2.5 - 3.5$ keV. The states $\ChiOne$ and $\ChiTwo$ interact with the standard model through a magnetic dipole operator \eqref{MDM}. As a result of this interaction, the dark matter $\ChiOne$ can get upscattered to $\ChiTwo$ somewhere in the Earth. The subsequent decay $\ChiTwo \rightarrow \ChiOne + \gamma$ can occur in a dark matter detector like DAMA, producing a  $\sim 2.5 - 3.5$ keV $\gamma$ ray that appears in the electronic channel of the detector. With the parameters chosen in section  \ref{Sec:DipolarDarkMatter}, this decay explains the observed signal at DAMA, since DAMA analyzes both nuclear and electron recoil events. 

Both aspects of this scenario, the initial upscattering process and the decays of the excited state, are constrained by observations. We begin with the initial, upscattering process which dumps energy in dark matter detectors. Following this, we analyze decays of the excited state.  We then briefly review collider and other astrophysical constraints on this scenario. 

\subsection{Upscattering}
\label{Sec:Upscattering}
The dark matter $\ChiOne$ interacts with nuclei through the operator  \eqref{MDM}. In this endothermic process, the kinetic energy $\sim \ChiMass v^2$ of $\ChiOne$ is partially consumed to provide the energy $\Splitting$ necessary to excite $\ChiOne \rightarrow \ChiTwo$. The remaining kinetic energy $\sim \ChiMass v^2 - \Splitting$ is shared between the kinetic energy of $\ChiTwo$ and the recoiling nucleus (of mass $m_N$). This scenario can explain the DAMA signal without running afoul of other experimental bounds only when $\ChiMass \lessapprox 1$ GeV (see sections  \ref{Sec:DipolarDarkMatter} and \ref{Sec:CDMS}). In this regime, $\ChiMass \ll m_N$ and the energy carried away by the recoiling nucleus is $\sim \frac{\ChiMass}{m_N} \Splitting \,  \lessapprox 0.10 \text{ keV } \l \frac{\text{28 GeV}}{m_N}\r \l \frac{\delta}{\text{3 keV}}\r \l \frac{m_{\chi}}{\text{1 GeV}}\r$. These recoils are below the thresholds of current dark matter experiments \cite{Graham:2010ca}. 

However, as pointed out in \cite{Graham:2010ca},  it is possible that a low threshold experiment like XENON 10 could nevertheless place a bound on such sub-threshold recoils due to upward fluctuations of photoelectron counts from such events.  Similarly, a low threshold, light nucleus experiment like CDMS Silicon could also have some sensitivity to such events, after accounting for detector efficiencies at such low recoil energies. The single nucleon upscattering cross-sections required to explain the DAMA events in this scenario are $\sim 10^{-36} - 10^{-38} \text{ cm}^2$ (see section \ref{Sec:DipolarDarkMatter}). These cross-sections were found to be safe from the above considerations for light dark matter in  \cite{Graham:2010ca}. In particular, the masses $\ChiMass \lessapprox 1$ GeV discussed in this paper are smaller than the dark matter masses considered in \cite{Graham:2010ca} and hence these events are even safer from such experimental constraints.

\subsection{Excited State Decay}
\label{Sec:Downscattering}
The decays of $\ChiTwo \rightarrow \ChiOne + \gamma$ produces $\sim 2.5 - 3.5$ keV $\gamma$ rays.  These $\gamma$ rays, which lie in the X-ray spectrum, can potentially be observed in detectors other than DAMA. Indeed, unlike a conventional WIMP which deposits energy in a detector due to scattering between the WIMP and the nuclei in the detector, the decay of $\ChiTwo$ and the subsequent $\gamma$ ray energy release occurs in vacuum. This process can therefore be observed not just in a conventional dark matter detector which measures  $\sim$ keV energy depositions (where it appears as pure ionization energy and gets classified as electron recoils), but also in any instrument which can observe events over a large {\it volume} even if the volume is a vacuum.  In this section, we discuss current experimental bounds on such decays. 

\subsubsection{XENON100}
\label{Sec:XENON100}
The XENON100 collaboration has released data from its preliminary run $\sim 11.17$ live days between October 20$^{th}$ to November 12$^{th}$, 2009 \cite{Aprile:2010um}. In this run, the experiment recorded $\sim 28$ events identified as electron recoils in the energy band between $\sim 5 - 33$ keVnr. The observations were performed in a fiducial volume that was a cylinder with radius $13.5$ cm and height $24.3$ cm. 

We now estimate the expected event rate in XENON100 as implied by the DAMA event rate. The actual event rate at DAMA is a function of the modulation fraction allowed by the underlying dark matter model.  With a large modulation rate, the actual number of dark matter events at DAMA is decreased resulting in better agreement between DAMA and other experiments. Consequently, we parameterize the XENON100 constraints on this model as the minimal modulation fraction necessary for the two event rates to be consistent. We note that large modulation fractions $\sim 50 - 70 \%$ are easy to obtain in this scenario since the initial excitation of $\ChiTwo$ requires inelastic, upscattering from $\ChiOne$  (see section \ref{Sec:DipolarDarkMatter}). In this scenario, the event rates at different experiments are related solely by the volumes of the two experiments. The other factor that affects the event rate is the time during which the observations were performed. XENON100 ran near the end of the month of October which is  close to the trough of the DAMA modulation. This factor must  be incorporated in the evaluation of the expected rate at XENON100. 

The $\sim 28$ events of XENON100 are spread over a wide energy region, between $\sim 5 - 33$ keVnr. In this model, the decays of $\ChiTwo$ produces a sharp X ray line at $\sim 3$ keV. But, this line is smeared by the energy resolution of XENON100. The  detector response at these low energies is very poorly understood \cite{Sorensen:2010hq}. The energy resolution is measured above $\sim 122$ keV and these measurements are used to extrapolate the resolution at low energies \cite{XENONResolution}. In this paper, as described in section \ref{Sec:DipolarDarkMatter}, we will use a Poisson distribution on the number of S1 photoelectrons which gives an energy resolution at  $\sim 3$ keV is roughly $\sim 1.14$ keV. With these parameters, the constraint imposed by XENON100 on this parameter space is evaluated and the results are shown in figure \ref{Fig:LDMPlot} (see section \ref{Sec:DipolarDarkMatter} for more details). These results indicate that the modulation fraction must be at least $\sim 50 \%$. 

%Incorporating the above and using an acceptance of $\sim 0.6$ for events in XENON100  \cite{Aprile:2010um} (this is the acceptance in the relevant energy bin $6 - 12$ keVnr at XENON100, see section \ref{Sec:DipolarDarkMatter}), we expect $\sim 14.5$ events during this run of XENON100. The expected number of events in this scenario is thus smaller than the total number observed at XENON100. 

%For this discussion, we assume that the modulation fraction is at least $\sim 60 \%$. In this case, the event rate at DAMA is $\sim 0.065$ events/day/kg/keV. Using this event rate, we can predict the rate at XENON100.

%With this resolution, the $\ChiTwo$ decays will cluster around $\sim 2 - 4$ keVee in XENON100. In the analysis \cite{Aprile:2010um} these events will appear between $\sim 6 - 12$ keVnr (since the nuclear recoils in XENON100 are quenched by $\sim 0.3$). The experiment observes $\sim 10$ events in this band while we expect $\sim 12.5$ events if the modulation fraction was higher than $70 \%$. Even at this level, the two rates are consistent, allowing for Poisson fluctuations of the event rate. 

We note that the limits derived in  figure \ref{Fig:LDMPlot} may be conservative. The parameter space available for the model can be significantly bigger owing to several experimental uncertainties. For example, the energy resolution at 3 keV could be significantly worse than the above estimate as it is based upon an extrapolation by nearly two orders of magnitude where measurements have not been made. In this case, the events would be spread out more evenly in XENON100 making this scenario more consistent. Consequently, the DAMA signal can be fit with lower modulation fractions, opening up the parameter space.

%For example, if the resolution was $\sim 1.6$ keV instead of $\sim 0.8$ keV, the $\ChiTwo$ decay events would be spread between $\sim 5  - 15$ keV where there are nearly $\sim 13$ events. Additionally, the event acceptance over this entire energy region is also uncertain \cite{Sorensen:2010hq}. If the acceptance was slightly smaller, then the event rate at XENON100 will also be correspondingly smaller. Consequently, the DAMA signal can be fit with lower modulation fractions, opening up the parameter space. 

In addition to ambiguities in the energy resolution of the detector, there is also uncertainty in the photoelectron yield of low energy $\sim 3$ keV X rays in XENON100. The analysis of  \cite{Aprile:2010um} was performed under the assumption that the photoelectron (p.e.) yield $L_y$ of Xenon was linearly proportional ($\propto 2.2 \l \frac{\text{p.e.}}{\text{keV}}\r$) to the X ray energy. However, these measurements were also made at $\sim 122.5$ keV \cite{Aprile:2010um}. It is unclear if this photoelectron yield remains linear at low energies. If $L_y$ was smaller, then the events would produce fewer photoelectrons and could appear in a region with a significantly smaller cut acceptance. In this case, there would be fewer expected events at XENON100, thereby opening up the parameter space of the model. 

%at energies below $\sim 5 - 15$ keV. In this region, the acceptance of the cuts is significantly smaller  \cite{Aprile:2010um} leading to fewer expected events. This would also open up the parameter space of this model that fits DAMA. 

\begin{figure}
\begin{center}
\includegraphics[width = 5.5 in]{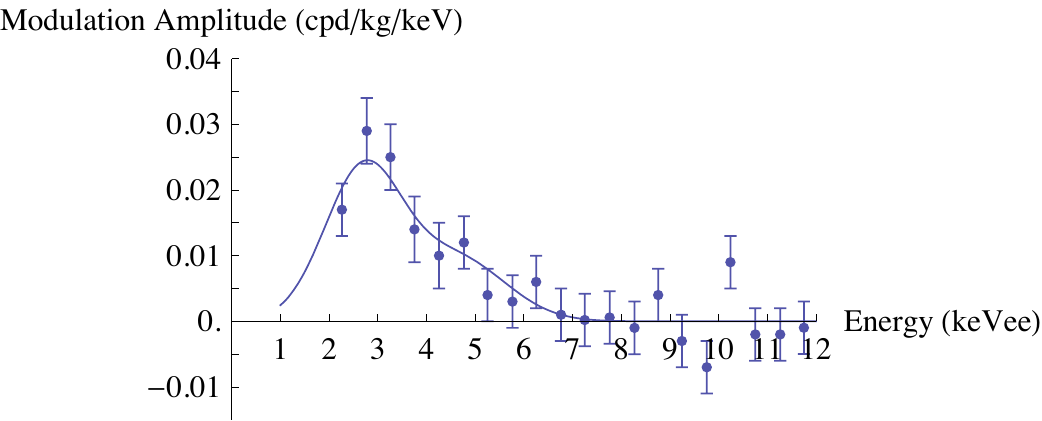}
\caption{ \label{Fig:DAMA2Lines} A comparison of the predicted and observed spectra at DAMA for two lines, one at 2.7 keV and the other at 4.7 keV.}    
\end{center}
\end{figure}

The above constraints were evaluated assuming that the signal in XENON100 was from a single $\sim 3$ keV line in the detector. However, the DAMA signal can also be fit through two spectral lines, one at around $\sim 2.7$ keV and another at $\sim 4.7$ keV (see figure \ref{Fig:DAMA2Lines}). Indeed, these two spectral lines are actually a better fit to the DAMA data than a single line.  In order to obtain two spectral lines instead of one, we need two excited states $\chi_{e_{1}}$ and $\chi_{e_{2}}$ that are split from the dark matter $\ChiOne$ by $\sim 2.7$ keV and $\sim 4.7$ keV respectively. The existence of such states may be generic in UV completions of this framework, such as in composite dark matter models   \cite{Alves:2009nf} and scenarios where the dark matter is charged under a broken non-Abelian group  \cite{ArkaniHamed:2008qn}. In this case, the events at XENON100 will not be clustered around $\sim 10$ keV, but would rather be more evenly spread across the region $\sim 5 - 25$ keVnr. Since there are significantly more events in this region, this also alleviates the constraints from XENON100 on this parameter space.

Despite these detector uncertainties, this model predicts that a significant fraction of  events observed at XENON100 between $\sim 5 - 25$ keVnr are due to dark matter. Consequently, this model predicts that XENON100 will continue seeing such electron recoil events in future runs of the experiment, and in fact even more during the summer. While the absence of such events will constrain this scenario, we note that the detector response at low energies has to be better understood before these constraints can be sharply imposed. 

\subsubsection{CDMS}
\label{Sec:CDMS}
The CDMS collaboration has recently analyzed its low energy electron recoil spectrum in its Germanium detector \cite{Ahmed:2009rh}. In order to facilitate a comparison between the event rates at CDMS and DAMA, this analysis assumed a $Z^2$ scaling in the  scattering cross-section. It was then found that the event rates in the two experiments would be compatible if the dark matter signal modulated at $\sim 12 \%$. In this scenario, the event rates at different experiments depends entirely on the volume of the detector. Comparing the relative volumes of the two detectors\cite{Ahmed:2009rh, Bernabei:2008yi}, we find that the event rate at CDMS is compatible with that of DAMA as long as the rate of dark matter induced electron recoils in DAMA is smaller than $\sim 0.09$ events/day/kg/keV. This implies that the observed annual modulation $\sim 0.022$ events/day/kg/keV in DAMA can be reproduced if the dark matter signal modulates at $\sim 24 \%$. This large modulation fraction is quite easy to achieve in the scenario discussed in this paper since the initial excitation of $\ChiTwo$ requires inelastic, upscattering from $\ChiOne$ (see section \ref{Sec:DipolarDarkMatter}).

We note that since these events are visible at CDMS, this scenario can be strongly constrained if the CDMS collaboration analyzes the annual modulation of its electron recoil events.  In this model, the annual modulation signal at DAMA implies a modulation of $\sim 0.015$ events/day/kg/keV at CDMS. The non-observation of such a modulation will rule out this scenario as an explanation for DAMA. 

\subsubsection{CoGeNT}
\label{Sec:CoGeNT}
The CoGeNT collaboration has released its observed recoil spectrum, consisting of both electron and nuclear recoil events \cite{Aalseth:2010vx}.  This experiment observed an event rate of at least $\sim  1.8$ events/kg/day/keVee (prior to efficiency corrections). Assuming at least a $24 \%$ modulation in the event rate at DAMA (see sub section \ref{Sec:CDMS})and scaling the expected event rates between the two experiments using the relative volume of the apparatus, we find that DAMA implies a rate $\sim 0.06$ events/kg/day/keVee at CoGeNT. This is of course smaller than the rate observed at CoGeNT and is hence not a constraint. However, with sufficient exposure, the 3 keVee line produced by the decays of $\ChiTwo$ may be observable at CoGeNT. Furthermore, just like CDMS, an analysis by CoGeNT of the annual modulation of its signal could constrain this scenario.

\subsubsection{CAST}
\label{Sec:CAST}
The decay of $\ChiTwo \rightarrow \ChiOne + \gamma$ occurs in vacuum and is visible in any volume that is instrumented to observe $\sim 2.5 - 3.5$ keV $\gamma$ rays. The CAST experiment \cite{Irastorza:2006gs}, constructed to search for solar axions, is an example of such an instrumented vacuum region that is sensitive to X rays.  The experiment consists of two parallel pipes of cross-sectional area $14.5 \text{ cm}^2$ and length $10$ m \cite{Irastorza:2006gs}. Detectors are placed at the ends of the pipes and are sensitive to X rays in the keV regime. CAST aims to look for solar axions by looking at the variation in the observed $\gamma$ ray count as the telescope's orientation with respect to the Sun is changed. In this scenario, the decays of $\ChiTwo$ gives rise to a constant background $\gamma$ flux at CAST. This background flux  is $\sim 4.32 \, \frac{\text{events}}{\text{cm}^2 \text{ day } \text{ keV}}$  \cite{Irastorza:2006gs}. Assuming at least a $24 \%$ annual modulation in DAMA (see section \ref{Sec:CDMS}), the event rate from $\ChiTwo$ decay is no greater than $\sim 3 \times 10^{-4} \, \frac{\text{events}}{\text{cm}^3 \text{ day } \text{ keV}} $ in the CAST volume. Accounting for the $\gamma$ rays that are oriented towards the detectors at the end of the telescope, we find that the expected flux from $\ChiTwo$ decays is $\lesssim 0.01 \, \frac{\text{events}}{\text{cm}^2 \text{ day } \text{ keV}}$, significantly smaller than the observed background. 

\subsubsection{Bubble Chamber Experiments}
\label{Sec:BubbleChamber}
Bubble chamber experiments such as  COUPP \cite{Behnke:2008zza} and PICASSO \cite{Archambault:2009sm} are capable of operating at low thresholds. In particular, COUPP has run with a threshold $\sim 5$ keV, while PICASSO has operated above $\sim 2$ keV. However, these experiments are intrinsically insensitive to electron and photon energy deposition below $\sim 350$ keV \cite{Behnke:2008zza,  Archambault:2009sm} and do not constrain the decays of $\ChiTwo$. Moreover, the initial upscattering of the dark matter $\ChiOne$ occurs at energies below $\sim 1$ keV for the parameter space considered in section \ref{Sec:DipolarDarkMatter}. This energy is below the threshold of these experiments and  is thus unconstrained. 

\subsubsection{Near Earth X Ray Emission}
\label{Sec:XRayTelescopes}
The dark matter $\ChiOne$ can be upscattered to $\ChiTwo$ from the Earth, the Moon and the Sun. The decays of the excited state $\ChiTwo$ produces $\gamma$ rays which can be detected by X-ray telescopes. The Sun is a powerful source of X-rays and the contribution of the dark matter to the solar X-ray flux is negligible in this scenario. Moreover, during the day,  the solar X ray flux illuminates the Earth and the Moon, resulting in significant X ray emission from them  \cite{Bhardwaj:2002xg}. However, at night, these objects are significantly darker in the X-ray spectrum and we will use X Ray measurements of this flux to constrain this scenario. We consider three kinds of measurements.

 First, satellites in low Earth orbit such as RTXE \cite{Revnivtsev:2003wm} and SWIFT \cite{Moretti:2008hs}, with orbits $\sim 600$ km have measured the cosmic X ray background from the dark side of the Earth. In these measurements, the satellite is behind the Earth, with its telescope facing the open cosmos. The telescope view is restricted to be at least $30^{\circ}$ away from the Earth horizon in order to avoid backgrounds from the sunlit portion of the Earth \cite{Moretti:2008hs}. These satellites are therefore sensitive to decays of $\ChiTwo$ that occur anywhere within the telescope field of view that lies beyond the orbit of the satellite. The decaying $\ChiTwo \rightarrow \ChiOne + \gamma$ produces a $\sim 3$ keV X ray line and the flux from these decays must be consistent with the measured cosmic X ray background.  Using the instrumental energy resolution $\sim 0.1$ keV of SWIFT, we require that the flux from the decaying $\ChiTwo$ at a $\sim 600$ km orbiting satellite be no greater than $\sim 0.03 \frac{\text{cts}}{\text{cm}^2 \text{ s sr} }$ in the energy bin around $\sim 3$ keV \cite{Moretti:2008hs}. 
 
 The flux of X rays at a local experiment from the decaying $\ChiTwo$ can be calculated as follows. The initial upscattering of the dark matter $\ChiOne$ occurs with a roughly uniform cross-section in the Earth, giving rise to a uniform density of $\ChiTwo$ inside the Earth. Away from the Earth, at a distance $r$ from the center of the Earth, the density  of $\ChiTwo$ drops $\propto \ r^{-2}$.  Ultimately, this density extends out to a distance $L \sim \frac{v_{f}}{\Gamma}$ set by the decay length of $\ChiTwo$ (see section \ref{Sec:DipolarDarkMatter}), after which the density is exponentially cut off. The X rays produced in $\ChiTwo$ decays in this entire region can strike the telescope and hence the total X ray flux is computed by integrating over the decays that occur in this entire region. 
 
 In this scenario, the density of $\ChiTwo$  inside the Earth is set by the event rate required to fit DAMA. The only other essential parameter that determines the flux in an orbiting telescope is the decay length of $\ChiTwo$. For a decay length significantly smaller than the Earth radius, the density of $\ChiTwo$ drops exponentially away from the Earth. In this case, the density of $\ChiTwo$ in the region that lies beyond a telescope that is on a $\sim 600$ km orbit is suppressed compared to the density of $\ChiTwo$ in DAMA, leading to a reduction in the X ray flux at the telescope. Performing the above calculations numerically, we find that the flux from the decays of  $\ChiTwo$ are smaller than the observed cosmic X ray background at an X ray telescope orbiting the Earth at $\sim 600$ km if the decay length of $\ChiTwo$ is smaller than $\sim 1000$ km (assuming that the DAMA modulation fraction is $\sim 50 \%$). With this choice of decay length, this scenario is consistent with the observations of the cosmic X ray background by  low Earth orbit satellites such as RTXE  \cite{Revnivtsev:2003wm} and SWIFT \cite{Moretti:2008hs}. 
  
Second, the RTXE satellite pointed its telescope at the dark side of the Earth and measured the X ray flux emerging from it. The collaboration assumed that the X ray flux from this region was entirely due to instrumental systematics, and subsequently subtracted this flux from its measurements of the clean X ray sky to get a systematics free measurement of the cosmic X ray background \cite{Revnivtsev:2003wm}. In this measurement, it was found that the X ray flux emerging from the dark side of the Earth was comparable to the cosmic X ray background.  Numerically evaluating this contribution, we find that the X ray flux at RTXE is consistent with  the measured background count rate in the $\sim 3$ keV bin as long as the decay length of $\ChiTwo$ is smaller than $\sim 900$ km. 

Finally, the Chandra X-ray telescope measured the flux of X-rays emerging from the dark side of the Moon \cite{Wargelin:2004vm}. This measurement was performed with the telescope pointed towards the dark side of the Moon. Chandra is therefore sensitive to the decays of $\ChiTwo$ that occur between its orbit and the lunar surface. There are two sources of $\ChiTwo$ in this region. First, upscattering from the Earth creates a local flux of $\ChiTwo$ in the region near the Earth. However, since the decay length of $\ChiTwo$ must be smaller than $\sim 900 - 1000$ km due to the preceding considerations, the flux of $\ChiTwo$ that lies beyond the perigee $\sim 16000$ km of Chandra is negligible and does not constrain this scenario. Second, the dark matter $\ChiOne$ can get upscattered in the Moon and then subsequently decay above the lunar surface. These X rays would then be visible to Chandra. However, with the parameters chosen in this model, the flux of X rays at Chandra produced by the dark side of the Moon are significantly smaller than the measurements of Chandra. 

In summary, near Earth measurements of the local X ray flux can impose significant constraints on the parameter space of this scenario. Consistency with these measurements requires that the decay length of the excited state $\ChiTwo$ be smaller than $\sim 1000$ km. This  is  the biggest constraint in achieving this scenario through an electric dipole moment (EDM) operator instead of the magnetic dipole operator \eqref{MDM} considered in this paper. In the case of an EDM operator, the scattering cross-section for $\ChiOne \rightarrow \ChiTwo$ is enhanced by $\frac{1}{v^2}$ \cite{Sigurdson:2004zp}. In order to match the event rate at DAMA, the initial upscattering rate can be made sufficiently small by taking the scale suppressing the EDM operator to be significantly higher than the scale used in \eqref{MDM}. But, in this case, the decay length of the dark matter also increases and in particular becomes larger than the Earth radius resulting in a conflict with the above X ray bounds. 

For the magnetic dipole moment operator \eqref{MDM} considered in this paper, there is a significant parameter space where the DAMA signal can be fit with the excited state decay length being shorter than $\sim 1000$ km.  The decay lengths are $\sim  100 - 1000$ km in the parameter space available to this model after the imposition of constraints from XENON100 (see section \ref{Sec:DipolarDarkMatter} and figure \ref{Fig:LDMPlot}). These decay lengths were computed using the $\ChiTwo$ lifetime \eqref{gamma} and integrating over a distribution of final state velocities of the $\ChiTwo$ produced after the upscattering of $\ChiOne$. We note that towards the lower end of this parameter space, when the decay lengths become smaller than $\sim 100$ km, these  X ray constraints entirely vanish since the decays are confined to the lower atmosphere which absorbs  $\sim 3$ keV X rays well before they can make it out to the orbits of X ray satellites.

\subsubsection{X Ray Line Emission}
\label{Sec:XRayLines}
Collisions with interstellar gas can upscatter the dark matter $\ChiOne$ to $\ChiTwo$  in the galaxy. The subsequent decay of $\ChiTwo$ produces a $\sim 3$ keV X ray line that  contributes to the galactic  X ray background. Line emissions from decaying dark matter in this energy band were searched for in \cite{Boyarsky:2006ag} and an upper bound $\sim 10^{27}$ s was placed on the dark matter lifetime for dark matter masses in the keV regime. For GeV dark matter producing keV lines, this bound is weakened to a lifetime of  $\sim 10^{21}$ s since the number density of a GeV mass dark matter particle is smaller than that of a keV dark matter particle by $10^{-6}$. In our scenario, this bound translates into a bound on the galactic upscattering rate. Most of the galaxy consists of hydrogen gas which has an  average number density of $\sim 1 \, \text{cm}^{-3}$ in the galactic disk. With this gas density, the upscattering rate is smaller than $10^{-21} \text{ s}^{-1}$ as long as the $\ChiOne$ upscattering cross-section is smaller than $\sim 10^{-28} \text{ cm}^{2}$. Similar bounds were also placed in \cite{Profumo:2006im}.

The upscattering cross-sections considered in this scenario are considerably smaller than the above limit. The typical scattering cross-section considered in this model is provided by \eqref{fullMDMsigma} which yields  per-nucleon scattering cross-sections $\sim 10^{-36} - 10^{-38} \text{ cm}^2$ in the Earth. But, this scattering cross-section is between the magnetic dipole of the dark matter $\ChiOne$ and the charge of the nucleus. As explained in section \ref{Sec:DipolarDarkMatter}, this is the relevant scattering cross-section in this scenario since most of the Earth is made up of nuclei that have no net spin. However, when there is a nuclear spin, there is an additional scattering process provided by the interaction between the magnetic dipole moments of the dark matter and the nucleus itself   \cite{Chang:2010en, Barger:2010gv}. For the case of hydrogen, this dipole-dipole scattering cross-section is larger by a factor of $\sim 4$   \cite{Chang:2010en, Barger:2010gv}. Consequently, this model would imply upscattering cross-sections $\sim 10^{-35} - 10^{-37} \text{ cm}^2$ in the galaxy between the dark matter and hydrogen, which is smaller than the limit imposed by  \cite{Boyarsky:2006ag}.

\subsubsection{Directional Detection of Dark Matter}
\label{Sec:DarkMatterDirectionDetection}
Significant progress has been recently achieved in constructing detectors that are sensitive to the direction of the incoming dark matter particles. Conventional inelastic dark matter models that cause nuclear energy deposition give rise to daily modulations in the direction of the dark matter signal in these experiments \cite{Finkbeiner:2009ug}. But,  in this case, the decay of $\ChiTwo$ emits a photon isotropically and hence the direction of the dark matter signal will not modulate. 

These experiments can nevertheless serve as powerful probes of luminous dark matter since they are large volume detectors. \footnote{We thank Neal Weiner for pointing this out to us.} Current runs of these experiments do not constrain this scenario since they were operated at high thresholds (for example, NewAge \cite{Miuchi:2010hn} and DRIFT-2 \cite{Burgos:2007zz}). DMTPC \cite{Ahlen:2010ub} has also published results of its initial run, with its cuts optimized to search for events at high recoil energies. Luminous dark matter will give rise to visible signals in these experiments if they can be operated at a threshold below $\sim 3$ keV with reduced backgrounds at these energies.  

\subsubsection{Neutrino Detectors}
\label{Sec:NeutrinoDetectors}
Neutrino detectors like Borexino \cite{Borexino} and Super Kamiokande \cite{Desai:2004pq} are large volume detectors. However, their operating thresholds are close to $\sim 1$ MeV and hence these experiments are insensitive to luminous dark matter. 

 \subsection{Collider,  Astrophysical and CMB Constraints}
\label{Sec:OtherConstraints}
Collider and other precision particle physics limits on light dark matter particles with magnetic moments were considered by \cite{Sigurdson:2004zp}. The most stringent of these limits arise from direct production of dark matter particles in the Tevatron. As per this analysis, dark matter particles with magnetic moments $\sim 10^{-17}$ e-cm $\sim \l  \text{TeV}\r^{-1} $ are safe from collider constraints. This is roughly the size of the magnetic moments considered in this paper (see section \ref{Sec:DipolarDarkMatter}). Moreover, as pointed out in \cite{Sigurdson:2004zp}, magnetic moments larger than the above limit are also allowed if the magnetic moment operator falls apart at Tevatron energies. Indeed, if this operator is radiatively generated by weak scale $\mathcal{O} \l 100 \text{ GeV}\r$ particles (see section \ref{Sec:DipolarDarkMatter}), it would fall apart at the scale of those particles. For example, if the operator falls apart at LEP energies, the dipole moment has to be smaller than $\l 1 \text{ TeV} \r^{-1}$ \cite{Masso:2009mu}. The parameter space considered in this paper satisfies this constraint. 

Astrophysical limits on dark matter interaction cross-sections with nuclei have also been placed. The most significant of these constraints arise from the capture and subsequent annihilation of dark matter into neutrinos. These are then constrained by measurements at Super Kamiokande \cite{Desai:2004pq, Hooper:2008cf, Griest:1986yu, Gould:1987ju}. Similarly, it has also been pointed out that the capture of dark matter in white dwarfs located in globular clusters with dark matter densities greater than $\sim 1000$ times the mean galactic dark matter density could also place limits on the dark matter - nucleon interaction cross-section \cite{McCulloughWD, HooperWD}. However, both these bounds are model dependent and can be easily overcome with variations of the underlying particle physics model  \cite{Graham:2010ca}. A parameter space similar to the one considered in this paper was discussed in  \cite{Graham:2010ca} and we refer the reader to the discussion in \cite{Graham:2010ca} on the applicability of these bounds. 

The CMB also constrains light dark matter \cite{Galli:2009zc} that annihilates to electromagnetic and colored standard model particles with a thermal annihilation cross-section. In fact,  dark matter masses below $\sim 10$ GeV are in conflict with WMAP5 observations if the dark matter dominantly annihilates to the above final states. In this model, this constraint can be naturally evaded. The magnetic dipole moment operator \eqref{MDM} connects two different states $\ChiOne$ and $\ChiTwo$, but vanishes by fermion anti-commutation when the two states are identical. During freeze out, the states $\ChiOne$ and $\ChiTwo$ are present in the plasma since they have nearly identical masses. The operator \eqref{MDM} then causes annihilations between these states and the standard model. For the parameters chosen in section \ref{Sec:DipolarDarkMatter}, this cross-section is roughly the thermal relic cross-section necessary to generate a dark matter abundance of $\ChiOne$ and $\ChiTwo$ (see figure \ref{Fig:LDMPlot}). $\ChiTwo$ then decays rapidly (well before nucleosynthesis) to $\ChiOne$ through \eqref{MDM}, resulting in a relic population of $\ChiOne$. In order for  \eqref{MDM} to induce annihilations of $\ChiOne$, the state $\ChiTwo$ has to be integrated out of the theory. This results in a dimension 7 operator that mediates the annihilation of $\ChiOne$ into two photons, suppressed by two powers of the dipole moment scale $\Lambda$ and one power of the dark matter mass. These annihilation cross-sections are significantly suppressed compared to the thermal relic annihilation cross-section between $\ChiOne$ and $\ChiTwo$ since the latter are mediated by a dimension 5 operator suppressed by only one power of $\Lambda$. Consequently, at the effective theory level, this model is intrinsically safe from CMB constraints on light dark matter. UV completions of this operator must however not open significant annihilation channels into  charged or colored standard model particles.

\section{Conclusions}

We have proposed a novel explanation for DAMA arising from the excitation of dark matter in the Earth followed by the decay of the excited state through single photon emission.  When the decay occurs inside a direct detection experiment it appears as electromagnetic, not nuclear, energy deposition.  This naturally avoids many of the constraints from direct detection experiments.  Both the upscattering and the decay, as well as possibly the relic abundance, are set by a single operator, the dipole moment operator.  Thus it is a tightly determined model from the beginning, with essentially only two free parameters, the dark matter mass and the scale suppressing the dipole moment operator.  Interestingly, for reasonable choices of both parameters, a mass $\sim 1$ GeV and a scale $\sim 1$ TeV, this model can explain DAMA and avoid all other constraints.  In particular, this model naturally has a higher annihilation cross section in the early universe when both states exist and a much lower cross section after the excited state population has decayed.  Thus the annihilation cross section drops significantly after freezeout, allowing the model to have the correct relic abundance but avoid the tight CMB constraints on light dark matter.

Unlike every other direct detection model, in this case the actual detector material is irrelevant.  The observed rate of events will only be proportional to the volume of the detector, not its mass or composition.  Experiments such as CDMS and XENON100 could see a signal of this model if they have sensitivity to low energy, $\lesssim 3$ keV, electronic events.  In fact, current results from XENON100 provide one of the strongest constraints on this model.  These experiments should see an annual modulation precisely proportional to DAMA's signal and the ratio of the volumes of the two experiments.  Such a signal should be clearly visible at XENON100 and CDMS if they measure the annual modulation of their electronic events.  Note that in this case a directional detection experiment will actually not reveal additional information since the decay of the dark matter emits the photon isotropically.  On the other hand, because the cross section for the original upscattering of the dark matter in the Earth does have angular dependence, there can be an observable daily modulation of the signal.  As the angle of the dark matter wind hitting the Earth near DAMA changes over the course of the day, the rate DAMA sees will also vary.  Given the high statistical significance of DAMA's signal, this daily modulation may also be visible if DAMA were to search for it in their data.  Over most of our parameter space however, this daily modulation is suppressed because the initial upscattering of the dark matter off a nucleus in the Earth is not strongly peaked in the forward direction.  This is partly because the dark matter is lighter than the nucleus and so can easily backscatter.  Nevertheless, this may make an interesting signal to search for.

\section*{Acknowledgments}

We would like to thank Sergei Dubovsky,  Roni Harnik, Kiel Howe,  Dan McKinsey,   Kaixuan Ni, Prashant Saraswat, Martin Schmaltz, Peter Sorensen, Jesse Thaler and Neal Weiner for useful discussions.  B.F.  is supported by DOE grant DE-FG02-01ER-40676. S.R. is supported by the DOE Office of Nuclear Physics under grant DE-FG02-94ER40818.  S.R. is also supported by NSF grant PHY-0600465.

\end{document}